\documentstyle[12pt]{article}
\renewcommand{\theequation}{\arabic{section}.\arabic{equation}}

\begin{document}
\date{}

\large{
\begin{center}
{\bf General Formulation of Quantum Analysis } \end{center} }

\vspace{0.3cm}

\begin{center}
Masuo SUZUKI
\end{center}

\vspace{0.3cm}

\begin{center}
Department of Applied Physics, Science University of Tokyo, \\
1-3, Kagurazaka, Shinjuku-ku, Tokyo 162, Japan 
\end{center}

\vspace{0.5cm}

\noindent {\bf Abstract}

A general formulation of noncommutative or quantum derivatives
for operators in
a Banach space is given on the basis of the Leibniz rule,
irrespective of their explicit representations such as the G\^ateaux
derivative or commutators. This yields a unified formulation of quantum
analysis, namely the invariance of quantum
derivatives, which are expressed by multiple integrals of ordinary
higher derivatives with hyperoperator variables.
Multivariate quantum analysis is also formulated in the present
unified scheme by introducing a partial inner derivation and a
rearrangement formula. Operator Taylor expansion formulas are also
given by introducing the two hyperoperators $
\delta_{A \rightarrow B} \equiv -\delta_A^{-1} \delta_B$ and $d_{A
\rightarrow B} \equiv \delta_{(-\delta_A^{-1}B) ; A}$ with the inner
derivation $\delta_A : Q \mapsto [A,Q] \equiv AQ-QA$. Physically
the present noncommutative derivatives
express quantum fluctuations and responses.

\vspace{0.5cm}
\noindent
{\bf I. Introduction}
\setcounter{section}{1}
\setcounter{equation}{0}

Recently noncommutative calculus has attracted the interest of many
mathematians and physicists$^{1-15}$. The present author$^{10-15}$
has introduced the quantum derivative $df(A)/dA$ of the operator
function $f(A)$ in the G\^ateaux differential$^{1-3}$
\begin{equation}
df(A)=\lim_{h \rightarrow 0} \frac{f(A+hdA)-f(A)}{h} \equiv
\frac{df(A)}{dA} \cdot dA. \end{equation}
Here the quantum derivative $df(A)/dA$ is a hyperoperator$^{10-15}$,
which maps an arbitrary operator $dA$ to the differential $df(A)$ in
a Banach space. There is also an algebraic definition$^{8,9,12,13}$
of the differential $df(A)$ as \begin{equation}
df(A)=[H, f(A)]
\end{equation}
for an auxiliary operator $H$ in a Banach space. This differential
depends on $H$. In particular, we have \begin{equation}
dA=[H,A].
\end{equation}
The property that $d^2A=0$ requires the following condition
\begin{equation}
[H,[H,A]]=[H, dA]=0.
\end{equation}

In the previous papers$^{10-13}$, we have shown that the differential
$df(A)/dA$ defined in (1.1) is expressed by \begin{equation}
\frac{df(A)}{dA}=\frac{\delta_{f(A)}}{\delta_A}, \end{equation}
where $\delta_A$ denotes an inner derivation defined by
\begin{equation}
\delta_AQ=[A,Q]=AQ-QA
\end{equation}
for an arbitrary operator $Q$ in a Banach space. The ratio of the two
hyperoperators in (1.5) is well defined$^{10-13}$ when $f(A)$ is a
convergent operator power series.

On the other hand, the derivative $df(A)/dA$ defined through the
commutator (1.2) is also expressed$^{9,12,13,16}$ by Eq.(1.5). This
is easily derived as follows. From Eq.(1.2), we have \begin{eqnarray}
\lefteqn{df(A)=\delta_H f(A)= -\delta_{f(A)} H= -\delta^{-1}_A
\delta_A \delta_{f(A)} H}\nonumber\\ & &
=\delta_A^{-1}\delta_{f(A)}(-\delta_A H)= \delta^{-1}_A \delta_{f(A)}
[H,A]= \frac{\delta_{f(A)}}{\delta_A} dA, \end{eqnarray}
using the commutativity of $\delta_A$ and $\delta_{f(A)}$. The
meaning of the formal inverse $\delta_A^{-1}$ in Eq.(1.7) will be discussed in
the succeeding section. The above results suggest that the quantum
derivative $df(A)/dA$ defined in Eq.(1.8) is invariant for any choice of
definitions of the differential $df(A)$. One of the main purposes of
the present paper
is to make a unified formulation of quantum analysis and to prove the
invariance of the quantum derivative $df(A)/dA$ defined in
\begin{equation}
df(A) \equiv \frac{df(A)}{dA} \cdot dA
\end{equation}
for any differential $df(A)$ satisfying the Leibniz rule
\begin{equation}
d(fg)=(df)g + f(dg).
\end{equation}

In Sec.II, some mathematical preparations are made on the formal
inverse $\delta_A^{-1}$ of the inner derivation $\delta_A$. In Sec.III
we present a general formulation of quantum derivatives using the
hyperoperators $\delta^{-1}_{A} \delta_{dA}$ and
$\delta_{(-\delta_A^{-1}dA) ;A}$. Theorem I states the invariance of
the differential $df(A)$ for any choice of definitions of $df(A)$.
Theorem II gives the invariance of the derivative,
$df(A)/dA$. Theorem III presents algebraic expressions of
higher differentials $ \{ d ^n f(A) \} $. Theorem IV gives
multiple integral representations of higher derivatives
$ \{ d^n f(A) / dA^n \} $. Theorem V presents a general
Taylor expansion formula of $ f (A + xB)$ in terms of higher
derivatives $ \{ d^n f(A)/ dA^n \}$ for the noncommutative operators
$A$ and $B$.
A shift-hyperoperator
${\cal S}_A(B) : f(A) \mapsto f(A+B)$ is also formulated. A general
formulation of
multivariate quantum analysis is given in Sec.IV, by introducing a
partial inner derivation and a rearrangement formula. In Sec.V, an
auxiliary operator method is briefly discussed, and it is extended to
multivariate operator functions. In Sec.VI, some general remarks and
applications to exponential product formulas are briefly mentioned.
Summary and discussion are given in Sec.VII.

\vspace*{0.5cm}
\noindent
{\bf II. Inner Derivation, its Formal Inverse and Uniqueness}
\setcounter{section}{2}
\setcounter{equation}{0}

In the present section, we introduce the two hyperoperators
$(-\delta_A^{-1} \delta_B)$ and $\delta_{(-\delta_A^{-1}B)}$, and
discuss the existence and uniqueness of these hyperoperators in the
domain ${\cal D}_A$, which is defined by the set of convergent power
series of the operator $A$ in a Banach space. In general, it seems to
be meaningless to use the symbol $\delta_A^{-1}$, because the inverse
of the inner derivation $\delta_A$ does not necessarily exist and
furthermore is not unique even if it exists. Fortunately in our
problem, only the combinations $(-\delta_A^{-1} \delta_B)$ and
$\delta_{(-\delta_A^{-1} B)}$ appear in our quantum analysis of
single-variable functions. Thus there is a possibility to define them
uniquely.

\vspace*{0.5cm}
\noindent
(i)  Hyperoperator $\delta_{A \rightarrow B} \equiv -\delta^{-1}_A
\delta_B$

First we show that the hyperoperator $(-\delta^{-1}_A \delta_B)$ is
well defined when it operates on a function $f(A)$ in the domain
${\cal D}_A$. For this purpose, we confirm that \begin{equation}
\delta_{A \rightarrow B} A \equiv (-\delta^{-1}_A \delta_B)A=
\delta^{-1}_A (-\delta_B A)=
\delta^{-1}_A \delta_A B=B,
\end{equation}
namely $\delta_{A \rightarrow B} : A \mapsto B$. More generally, we
have \begin{equation}
\delta_{A \rightarrow B} A^n = \sum^{n-1}_{k=0} A^k(\delta_{A
\rightarrow B} A)A^{n-k-1} = \sum^{n-1}_{k=0} A^kBA^{n-k-1}
\end{equation}
for any positive integer $n$. Thus, the hyperoperator $\delta_{A
\rightarrow B} \equiv -\delta^{-1}_A \delta_B$ is well defined, at
least,
in the domain ${\cal D}_A$. Thus, the existence of $\delta_{A
\rightarrow B}$ has been shown, but it is not unique. In fact, we put
\begin{equation}
\delta_{A \rightarrow B} f(A)=F(A,B)
\end{equation}
which is constructed by the above procedure. Then, $F(A,B)+G(A)$ may
be also a solution of $(-\delta^{-1}_A \delta_B)f(A)$,
because\begin{equation} -\delta_B f(A)=\delta_A F(A,B)+ \delta_AG(A)
\end{equation}
for any operator $G(A)$ in a Banach space. If we impose, besides the
Leibniz rule,  the linearity of the
hyperoperator $\delta_{A \rightarrow B}$, namely
\begin{equation}
\delta_{A \rightarrow B} (f(A)+g(A))=
\delta_{A \rightarrow B}f(A) +\delta_{A \rightarrow B} g(A),
\end{equation}
and
\begin{equation}
\delta_{A \rightarrow B} (af(A))=a \delta_{A \rightarrow B} f(A)
\end{equation}
for a complex number $a$, then the uniqueness of $\delta_{A \rightarrow B}$
is assured. In fact, the expression $F(A,B)$ in (2.3) is obtained
explicitly by using this linearity of the hyperoperator $\delta_{A
\rightarrow B}$.

In order to study the role of the hyperoperator $\delta_{A
\rightarrow B}$ more explicitly, we introduce the symmetrized product
$\{A^m B^n\}_{{\rm {sym}}(A,B)}$ by \begin{equation} \{A^m
B^n\}_{{\rm {sym}}(A,B)} \equiv \sum_{k_1+ \cdots + k_{n+1}=m, k_j
\geq 0} A^{k_1}B A^{k_2} \cdots A^{k_n}B A^{k_{n+1}},
\end{equation}
where $m,n, \{k_j\}$ denote non-negative integers. This
symmetrized product is also written as
\begin{equation}
\{A^m B^n\}_{{\rm {sym}}(A,B)}=\frac{1}{n!} \left[\frac{d^n}{dx^n}
(A+xB)^{m+n} \right]_{x=0}. \end{equation}
Then, Eq.(2.2) is expressed by
\begin{equation}
\delta_{A \rightarrow B} A^m = \{A^{m-1} B\}_{{\rm {sym}}(A,B)}.
\end{equation}

Hereafter, we write $\{\cdots\}_{{\rm {sym}}(A,B)}$ simply as
$\{\cdots\}_{\rm {sym}}$, when no confusion arises. Similarly we
obtain
\begin{equation}
\delta_{A \rightarrow B} \{A^{m} B\}_{\rm {sym}} =\{A^{m-1}
B^2\}_{\rm {sym}},
\end{equation}
because
\begin{eqnarray}
\lefteqn{
-\delta_B \{A^m B\}_{\rm {sym}}=-\delta_B (\sum^m_{k=0} A^k B
A^{m-k})}\nonumber\\ & & =\sum^m_{k=0}[(-\delta_B A^k) B A^{m-k}+
A^kB (-\delta_B A^{m-k})] \end{eqnarray}
using the Leibniz rule. Using the commutativity of $A$ and $\delta_A$
and the relation (2.9), namely $-\delta_B A^m=
\delta_A \{A^{m-1} B\}_{\rm {sym}}$, we have \begin{eqnarray}
-\delta_B \{A^{m} B\}_{\rm {sym}}& = &
\sum^m_{k=1}(\delta_A \{A^{k-1} B\}_{\rm {sym}}) BA^{m-k} +
\sum^{m-1}_{k=0} A^kB
(\delta_A \{A^{m-k-1} B\}_{\rm {sym}})\nonumber\\ & = & \delta_A
\{A^{m-1} B^2\}_{\rm {sym}}. \end{eqnarray}
In general, we have the following formula.

\vspace*{0.3cm}
\noindent
{\bf Formula 1} : For non-negative integers $m (\geq 1)$ and $n$ and
for any operators $A$ and $B$ in a Banach space, we have
\begin{equation}
-\delta_B \{A^{m} B^n\}_{\rm {sym}} =\delta_A \{A^{m-1}
B^{n+1}\}_{\rm {sym}} \end{equation}
namely
\begin{equation}
\delta_{A \rightarrow B} \{A^{m} B^n\}_{\rm {sym}} = \{A^{m-1}
B^{n+1}\}_{\rm {sym}}
\end{equation}
Consequently, the domain of the hyperoperator $\delta_{A \rightarrow
B}$ is extended to the region ${\cal D}_{{\rm {sym}}(A,B)}$ which is
a set of convergent noncommuting symmetrized power series of $A$ and
$B$.

The proof of this formula is given as follows. First note that
\begin{equation}
\delta_{A+xB}(A+xB)^{m+n}=0,
\end{equation}
namely
\begin{equation}
-x \delta_B(A+xB)^{m+n}=\delta_A(A+xB)^{m+n}. \end{equation}
By comparing the $(n+1)$-th terms of the both sides of (2.16) in $x$
and using the relation (2.8), we obtain Eq.(2.13) and consequently
Eq.(2.14). An alternative derivation of Eq.(2.13) will be given by
extending the procedure shown in Eqs.(2.11) and (2.12).

Next we study the property of the power hyperoperators $\{\delta_{A
\rightarrow B}^k \}$. It is easy to show the following formula.

\vspace*{0.3cm}
\noindent
{\bf Formula 2} : For non-negative integers $k, m ( \geq k )$ and $n$
and for any operators in a Banach space, we have
$$
\delta_{A \rightarrow B}^k \{A^{m} B^n\}_{\rm {sym}} = \{A^{m-k}
B^{n+k}\}_{\rm {sym}}, \eqno{(2.17{\rm a})}
$$
and
$$
\delta^k_{A \rightarrow B} \{ A^m B^n \}_{\rm {sym}}
= 0 \;\;\;\;\; {\rm {if}} \;\;\;\;\; m<k. \eqno{(2.17{\rm b})}
$$
This gives the following general formula.

\vspace*{0.3cm}
\noindent
{\bf Formula 3} : When $f(A)$ is a convergent operator power series
of an operator $A$ in a Banach space, we have \addtocounter{equation}{1}
\begin{eqnarray}
\delta_{A \rightarrow B}^n \{f(A) B^m\}_{\rm {sym}} & = &
\int^1_0dt_1 \int^{t_1}_0dt_2 \cdots \int^{t_{n-1}}_0dt_n
\{f^{(n)}(t_nA) B^{m+n}\}_{\rm {sym}}\nonumber\\ & = &
\frac{1}{(n-1)!} \int^1_0 dt (1-t)^{n-1} \{f^{(n)}(tA) B^{m+n}\}_{\rm
{sym}}.
\end{eqnarray}
Here, $f^{(n)}(x)$ denotes the $n$th derivative of $f(x)$.

\vspace*{0.5cm}
\noindent
(ii)  Hyperoperators $\delta_{(-\delta_A^{-1}B)}$ and $d_{A \rightarrow
B} \equiv \delta_{(-\delta_A^{-1} B) ; A}$

An operator $H$ defined by
\begin{equation}
-\delta_A H = B
\end{equation}
does not necessarily exist, as is well known. However, the
hyperoperator $\delta_{(-\delta_A^{-1}B)}$ is well defined, at least,
when it operates on $f(A)$ in the domain ${\cal D}_A$ for an operator
$A$ in a Banach space. In fact, we can interpret it as
\begin{eqnarray}
\delta_{(-\delta_A^{-1}B)} A^m& = & \sum^{m-1}_{k=0}
A^k(\delta_{(-\delta_A^{-1} B)}A) A^{m-k-1} = \sum^{m-1}_{k=0}
A^k(\delta_A \delta_A^{-1} B)A^{m-k-1}\nonumber\\
& = & \sum^{m-1}_{k=0} A^k B A^{m-k-1}
=\delta_{A \rightarrow B}A^m.
\end{eqnarray}
In other words, the formal hyperoperator $\delta_A^{-1}$ in
$\delta_{(-\delta_A^{-1}B)}$ should be interpreted as a hyperoperator
operating on the left-hand-side hyperoperator $\delta_A$ (not on the
right-hand-side operator B). In this interpretation, the
hyperoperator $\delta_{(-\delta_A^{-1}B)}$ is defined even when the
operator $(-\delta_A^{-1}B)$ does not exist.

In general, we obtain the following formula.

\vspace*{0.3cm}
\noindent
{\bf Formula 4}: Under the requirement of the linearity of the
hyperoperators $\delta_{A \rightarrow B}$ and
$\delta_{(-\delta_A^{-1}B)}$, we have \begin{equation}
\delta_{(-\delta_A^{-1}B)} f(A) =\delta_{A \rightarrow B}f(A)=
\int^1_0\{f^{(1)}(tA) B\}_{{\rm sym}}dt
\end{equation}
for any operator $f(A) \in {\cal D}_A$. Here, $f^{(1)}(x)$ denotes
the first derivative of $f(x)$.

It should be remarked that the hyperoperator
$\delta_{(-\delta_A^{-1}dA)}$ is a kind of differential defined only
in the domain ${\cal D}_A$, whereas the hyperoperator $\delta_{A
\rightarrow dA}$ is defined in a wider domain but is not a
differential in the domain ${\cal D}_{{\rm {sym}}(A,dA)}$ outside of
the domain ${\cal D}_A$.

As was discussed before, the operator $H \equiv -\delta_A^{-1}B$ does
not necessarily exist, and it is difficult to define the power
hyperoperators $\{\delta^n_{(-\delta_A^{-1}B)}\}$ for $n \geq 2$ when
$H \equiv -\delta_A^{-1}B$ does not exist. Furthermore, they are
complicated$^{12}$ even if they do exist, unless $H$ commutes with
$B$. Thus, we define the following partial inner derivation
\begin{equation}
d_{A \rightarrow B} \equiv
\delta_{(-\delta_A^{-1}B) ;A},
\end{equation}
by which the commutator $\delta_{(-\delta_A^{-1}B)} $ is taken only
with the operator $A$ in a multivariate operator $f(A,B)$. For
example, we have \begin{equation}
d_{A \rightarrow B}(ABA)= (\delta_{(-\delta_A^{-1}B)} A)
BA+AB(\delta_{(-\delta_A^{-1}B)} A). \end{equation}
This new hyperoperator
$d_{A \rightarrow B}$ is defined in the domain ${\cal D}_{A,B}$ which
is a set of convergent noncommuting power series of the operators $A$
and $B$. Clearly,	$d_{A \rightarrow B}$ is a kind of differential
satisfying the Leibniz rule for $B=dA$.

Next we study the power hyperoperators $\{d^n_{A \rightarrow B}\}$.
Clearly they are also differentials defined in the domain ${\cal
D}_{A,B}$. It will be interesting to find the relation between
$d^n_{A \rightarrow B}$ and $\delta^n_{A \rightarrow B}$.

First note that
\begin{eqnarray}
\lefteqn{
d^2_{A \rightarrow B} A^m=d_{A \rightarrow B} \{A^{m-1}B\}_{{\rm
sym}} =
d_{A \rightarrow B} \left(\sum^{m-1}_{j=0} A^j  BA^{m-j-1}\right)}\nonumber\\
& & = \sum^{m-1}_{j=1} \sum^{j-1}_{k=0} A^{j-k-1} BA^k BA^{m-j-1}
+ \sum^{m-2}_{j=0} \sum^{m-j-2}_{k=0} A^j BA^k BA^{m-j-k-2}\nonumber\\
& & = 2\{A^{m-2}B^2\}_{{\rm sym}}=2 \delta^2_{A
\rightarrow B}A^m
\end{eqnarray}
for $m \geq 2$. In general, we obtain the following formula.

\vspace*{0.3cm}
\noindent
{\bf Formula 5} : For $m \geq n$ and for any operators $A$ and $B$ in
a Banach space, we have \begin{equation}
d^n_{A \rightarrow B} A^m= n ! \delta^n_{A \rightarrow B} A^m.
\end{equation}
We have also $d^n_{A \rightarrow B} A^m = 0$ for $m < n$.
More generally, we have
\begin{equation}
d^n_{A \rightarrow B} f(A)= n ! \delta^n_{A \rightarrow B} f(A),
\end{equation}
when $f(A) \in {\cal D}_A$.

The proof of Formula 5 is given by mathematical induction using
the following lemma and Formula 2.

\vspace*{0.3cm}
\noindent
{\bf Lemma 1} : For non-negative integers $m ( \geq 1)$ and $n$ and
for any operators $A$ and $B$ in a Banach space, we have
\begin{equation}
d_{A \rightarrow B} \{A^m B^n\}_{{\rm sym}} = (n+1)
\{A^{m-1}B^{n+1}\}_{{\rm sym}}. \end{equation}

This is easily proved by using the definition (2.7) of
$\{ A^m B^n \}_{\rm sym} $ as in Eq.(2.24).
Formula 5 can be also confirmed directly from
the consideration on the number of permutations of $B^n$. More
intuitively, $\delta^n_{A \rightarrow dA}$ denotes an ordered partial
differential$^{11}$. On the other hand, $d^n_{A \rightarrow dA}$
denotes the $n$th differential, as will be discussed later.
Consequently we have Formula 5.

It should be remarked here that the hyperoperator $d^n_{A \rightarrow
B}$ is equivalent to $\delta^n_{(-\delta^{-1}_{A} B)}$ when $H \equiv
-\delta_A^{-1}B$ exists and it commutes with $B$. This equivalence
has been already used implicitly in the previous papers$^{12,13}$.

With these preparations, we discuss a general theory of derivatives
of $f(A)$ with respect to the operator $A$ itself in the succeeding
section.
\newpage
\noindent
{\bf III. Quantum Derivative, its Invariance and Operator Taylor
Expansion}
\setcounter{section}{3}
\setcounter{equation}{0}

In the present section, we give a general formulation of quantum
derivatives $\{d^nf(A)/dA^n\}$ which do not depend on the definition
of the differential $df(A)$. Our starting point of this general
theory is that the differential hyperoperator ``$d$" satisfies the
Leibniz rule (1.9) and that it is a linear hyperoperator.

\vspace*{0.3cm}
\noindent
(i) Quantum derivative and its invariance

Now we start with the following identity
\begin{equation}
Af(A)=f(A)A,
\end{equation}
when $f(A) \in {\cal D}_A$. Then, we have
\begin{equation}
d(Af(A))=d(f(A)A),
\end{equation}
which is rewritten as
\begin{equation}
(dA)f(A) + Adf(A)=(df(A))A + f(A)dA,
\end{equation}
using the Leibniz rule. This is rearranged as follows :
\begin{equation}
Adf(A)-(d f(A))A= f(A)dA-(dA)f(A).
\end{equation}
That is, we have
\begin{equation}
\delta_A df(A)=\delta_{f(A)} dA.
\end{equation}
This is our desired formula on the differential $df(A)$.

In order to discuss the solution of Eq.(3.5), we rewrite Eq.(3.5) as
\begin{equation}
\delta_Ad f(A) = -\delta_{dA}f(A).
\end{equation}
Obviously, $df(A)$ has a linearity property with respect to $f(A)$. Thus,
the solution $df(A)$ of Eq.(3.6) is uniquely given in the form
\begin{equation}
df(A)=\delta_{A \rightarrow dA}f(A),
\end{equation}
using the hyperoperator
$\delta_{A \rightarrow dA} \equiv -\delta_A^{-1} \delta_{dA}$ introduced
in Section II. This is also rewritten as
\begin{eqnarray}
df(A) & = & \int^1_0 dt\{f^{(1)}(tA)dA\}_{{\rm {sym}}(A,dA)}\nonumber\\ &
= & \int^1_0 f^{(1)}(A- t\delta_A) dt \cdot dA,
\end{eqnarray}
using Formula 4, namely Eq.(2.21).

The second equality of Eq.(3.8) is
proven as follows. First we prove it when $f(A)=A^m$ for an arbitrary
positive integer $m$. Clearly we have \begin{eqnarray}
dA^m & = & \int^1_0 dt\{f^{(1)}(tA)dA\}_{{\rm {sym}}(A,dA)}\nonumber\\ & =
&(m \int^1_0 t^{m-1}dt)\{A^{m-1}dA\}_{{\rm {sym}}(A,dA)}\nonumber\\ & = &
\{A^{m-1}dA\}_{\rm {sym}}.
\end{eqnarray}
On the other hand, we obtain
\begin{eqnarray}
\lefteqn{
\int^1_0 f^{(1)}(A- t\delta_A)dt \cdot dA =\int^1_0
f^{(1)}((1-t)A+t(A-\delta_A))dt \cdot dA}\nonumber\\ & & =m
\sum^{m-1}_{k=0} \;_{m-1}C_k
\int^1_0 (1-t)^k t^{m-1-k}dtA^k(A-\delta_A)^{m-1-k} \cdot dA\nonumber\\ &
& =m \sum^{m-1}_{k=0} \;_{m-1}C_k B(k+1, m-k) A^k(A-\delta_A)^{m-1-k}
\cdot dA\nonumber\\ & & =\sum^{m-1}_{k=0} A^k(A-\delta_A)^{m-1-k} \cdot
dA\nonumber\\ & & =\sum^{m-1}_{k=0} A^k(dA)A^{m-1-k} = \{A^{m-1}dA\}_{{\rm
sym}}, \end{eqnarray}
using the beta function $B(x,y)$, the binomial coefficient $\;_mC_k$, the
commutativity of $A$ and $\delta_A$,
and the following relation$^{10}$
\begin{equation}
(A- \delta_A)^n \cdot dA=(dA) \cdot A^n. \end{equation}
Thus, the second equality of Eq.(3.8) holds for $f(A) \in {\cal D}_A$.

Furthermore we can derive the following relation.

\vspace*{0.3cm}
\noindent
{\bf Lemma 2} : When $f(A) \in {\cal D}_A$, we have
\begin{equation}
\delta_{f(A)}=f(A)-f(A- \delta_A).
\end{equation}

Using this lemma, we obtain
\begin{eqnarray}
\delta_{f(A)}& = & f(A) -f(A-\delta_A)\nonumber\\
& = &\delta_A(f^{(1)}(A) - \frac{1}{2} \delta_A f^{(2)}(A) +
\cdots + \frac{(-1)^{n-1}}{n !} \delta_A^{n-1} f^{(n)}(A) + \cdots
)\nonumber\\
& = & \delta_A \int^1_0 f^{(1)}(A-t\delta_A)dt. \end{eqnarray}
This is formally written as
\begin{equation}
\int^1_0 f^{(1)} (A-t\delta_A)dt
=\delta_A^{-1} \delta_{f(A)}
=\frac{\delta_{f(A)}}{\delta_A}.
\end{equation}

Thus, summarizing Eqs.(3.5),(3.7),(3.8) and (3.14),
we obtain the following theorem on the differenial
$df(A)$.

\vspace*{0.3cm}
\noindent
{\bf Theorem I} : {\it When} $f(A) \in {\cal D}_A$, {\it we have}
\begin{equation}
\delta_A df(A)= \delta_{f(A)}dA,
\end{equation}
{\it and consequently}
\begin{eqnarray}
df(A)& = & \delta_{A \rightarrow dA} f(A)\nonumber\\ &=& \int^1_0
f^{(1)}(A-t \delta_A)dt \cdot dA= \frac{\delta_{f(A)}}{\delta_A} \cdot dA
\end{eqnarray}
{\it for any choice of definitions of the differential} $df(A)$.

It should be noted that the ratio of the two hyperoperators
$\delta_{f(A)}$ and $\delta_A$ is well defined for $f(A) \in {\cal D}_A$,
as was discussed in the preceding section. We define$^{10-12}$
the quantum derivative $df(A)/dA$ in Eq.(1.8), namely
\begin{equation}
df(A)=\frac{df(A)}{dA} \cdot dA.
\end{equation}
That is, the derivative $df(A)/dA$ is a hyperoperator which maps an
arbitrary operator $dA$ to the differential $df(A)$ given by Eq.(3.16).
Thus, we arrive at the following invariance theorem on the quantum
derivative defined in Eq.(3.17).

\vspace*{0.3cm}
\noindent
{\bf Theorem II (Invariance of Quantum Derivative)} : {\it When} $f(A) \in
{\cal D}_A$, {\it the quantum derivative} $df(A)/dA$ {\it is invariant for
any choice of definitions of the differential} $df(A)$ {\it satisfying the
Leibniz rule,} {\it and it is given by}
\begin{equation}
\frac{df(A)}{dA}=\frac{\delta_{f(A)}}{\delta_A}= \int^1_0 f^{(1)}(A-t
\delta_A)dt.
\end{equation}

Clearly, the ratio of the two hyperoperators $\delta_{f(A)}$ and
$\delta_A$ does not depend on the choice of definitions of the
differential $df(A)$. This invariance has been also discussed by
Nazaikinskii et al$^{9}$. in a different formulation based on the Feynman
index method. The present confirmation of the invariance is more direct
and transparent.

\vspace*{0.3cm}
\noindent
(ii) Higher derivatives and operator Taylor expansion

Now we discuss higher-order differentials $\{d^nf(A)\}$ and higher
derivatives\\ $\{d^nf(A)/dA^n\}$.

\vspace*{0.3cm}
\noindent
(ii-1) Higher-order differentials and derivatives

The hyperoperator $d_{A \rightarrow B}$ introduced in Eq.(2.22) is a
derivation satisfying the Leibniz rule (1.9). Thus, $d_{A \rightarrow B}$
is a kind of differential hyperoperator, when $B=dA$. We prove here the
following theorem.

\vspace*{0.3cm}
\noindent
{\bf Theorem III}: {\it The $n$th differential $d^nf(A)$ is expressed by}
\begin{equation}
d^nf(A)=d^n_{A \rightarrow dA}f(A)
\end{equation}
{\it for any choice of definitions of the differential} $df(A)$, {\it
when} $f(A) \in {\cal D}_A$.

The proof is given as follows. First note the following recursive
formula$^{10}$(3.21) obtained by differentiating Eq.(3.5), namely
\begin{equation}
\delta_A df(A)=\delta_{f(A)} dA
\end{equation}
repeatedly.

\vspace*{0.3cm}
\noindent
{\bf Formula 6} : When $f(A) \in {\cal D}_A$, we have \begin{equation}
\delta_A d^n f(A)=n \delta_{d^{n-1}f(A)} dA=-n \delta_{dA} d^{n-1}f(A).
\end{equation}

This gives the following result :

\vspace*{0.3cm}
\noindent
{\bf Formula 7} : When $f(A) \in {\cal D}_A$, we have \begin{equation}
d^n f(A)=n \delta_{A \rightarrow dA} d^{n-1}f(A)=n! \delta^n_{A
\rightarrow dA} f(A), \end{equation}
using the hyperoperator $\delta_{A \rightarrow B}=-\delta^{-1}_A \delta_B$
introduced in Sec.II.

Here we have also used the relation $df(A)=\delta_{A \rightarrow dA} f(A)$
given in Eq.(3.7). Using Formula 5, namely Eq.(2.26), we arrive at Theorem
III. This result means that any differential hyperoperator $d$ is
generally expressed by \begin{equation}
d=d_{A \rightarrow dA}
\end{equation}
in the domain ${\cal D}_{A,dA}$.

Next we define$^{10}$ the higher derivatives $\{d^nf(A)/dA^n\}$ through
the relation
\begin{equation}
d^n f(A)=\frac{d^nf(A)}{dA^n} : \underbrace{dA \cdot {\mbox {- - - }}
\cdot dA}_n. \end{equation}
Here, $d^nf(A)/dA^n$ denotes a hyperoperator which maps a set of the
operators $
(dA, \cdots , dA) \equiv : dA \cdot {\mbox {- - - }} \cdot dA \; \; {\rm
to} \; \; d^nf(A).
$
In an ordinary mathematical notation, one may prefer to write as
\begin{equation}
d^n f(A)=\frac{d^nf(A)}{dA^n} (\underbrace{dA, \cdots, dA}_n).
\end{equation}
However, as was emphasized before$^{10}$, the product form (3.24) is
essential in the present quantum analysis. That is, we use the product
form (3.24) only when the derivative $d^nf(A)/dA^n$ is expressed
explicitly in terms of $A$ and the inner derivations $\{\delta_j\}$
defined by$^{10}$ \begin{equation}
\delta_j :\underbrace{dA \cdot {\mbox {- - - }} \cdot dA}_n
=(dA)^{j-1}(\delta_AdA) (dA)^{n-j}. \end{equation}
If we use the notation (3.25), this property of product (3.26) and
\begin{equation}
A : \underbrace{dA \cdot {\mbox {- - - }} \cdot dA}_n =A(dA)^n
\end{equation}
can not be shown explicitly. Clearly $A$ and $\{\delta_j\}$ are commutable
with each other.

\vspace*{0.3cm}
\noindent
(ii-2) Integral representation of $d^nf(A)/dA^n$

Here we express $d^nf(A)/dA^n$ explicitly in an integral form in terms of
the above inner derivations $\{\delta_j\}$. Our result is given by the
following theorem.

\vspace*{0.3cm}
\noindent
{\bf Theorem IV} : {\it When} $f(x)$ {\it is analytic and} $f(A) \in {\cal
D}_A$, {\it any higher derivative} $d^nf(A)/dA^n$ {\it exists uniquely for
any choice of definitions of the differential} $df(A)$, {\it and it is
given explicitly in the form} \begin{equation}
\frac{d^nf(A)}{dA^n}=n! \int^1_0dt_1 \int^{t_1}_0dt_2 \cdots
\int^{t_{n-1}}_0dt_n f^{(n)}(A-\sum^n_{j=1}t_j \delta_j). \end{equation}
{\it Here,} $f^{(n)}(x)$ {\it denotes the} $n${\it th} {\it derivative of}
$f(x)$.

The proof is given as follows.
Once the above integral representation (3.28) is derived, the uniqueness
of it is clear. In the case of $n=1$, we have \begin{equation}
\frac{df(A)}{dA}=\frac{\delta_{f(A)}}{\delta_A}
=\frac{f(A)-f(A-\delta_A)}{\delta_A}=\int^1_0 f^{(1)}(A-t\delta_A)dt
\end{equation}
from Theorem II and Lemma 2.

The $n$th derivative of $f(A)$ divided by $n !$, namely $\hat{f_n}(A,
\delta_1, \cdots , \delta_n)$ defined by \begin{equation}
\hat{f_n}(A, \delta_1, \cdots , \delta_n) : (dA)^n \equiv \frac{1}{n !}
\frac{d^nf(A)}{dA^n} : (dA)^n =\delta^n_{A \rightarrow dA} f(A)
\end{equation}
is shown from Formula 6 to satisfy the following relation
$$
(\delta_1+ \cdots +\delta_n) \hat{f_n}(A, \delta_1, \cdots , \delta_n) =
\hat{f}_{n-1}(A, \delta_1, \cdots , \delta_{n-1})
-\hat{f}_{n-1}(A-\delta_1, \delta_2 \cdots , \delta_n). \eqno{(3.31{\rm a})}
$$
When $f(A) = A^k$, Eq.(3.31a) means that
$$
\delta_A d^n A^k = n (d^{n-1} A^k dA -d^{n-1} (A-\delta_A)^k dA),
\eqno{(3.31{\rm b})}
$$
which is equivalent by Formula 6 to saying that
$$
dAd^{n-1} A^k = d^{n-1} (A - \delta_A )^k dA.
\eqno{(3.31{\rm c})}
$$
The solution of (3.31a) with the condition (3.29) for $n=1$ is proven to be
given by
\addtocounter{equation}{1}
\begin{equation}
\hat{f_n}(A, \delta_1, \cdots , \delta_n)= \int^1_0dt_1 \int^{t_1}_0dt_2
\cdots
\int^{t_{n-1}}_0dt_n f^{(n)}(A-\sum^n_{j=1}t_j \delta_j),
\end{equation}
using the commutativity of $A$ and $\{\delta_j\}$, and the following
formula for $t=1$.

\vspace*{0.3cm}
\noindent
{\bf Formula 8} : For any positive integers $m$ and $n$, we have
\begin{eqnarray}
\lefteqn{(x_1+ \cdots + x_n) \int^t_0dt_1 \int^{t_1}_0dt_2 \cdots
\int^{t_{n-1}}_0dt_n f^{(m+1)}(tx-\sum^n_{j=1}t_j x_j)}\nonumber\\ & & =
\int^t_0dt_1 \int^{t_1}_0dt_2 \cdots \int^{t_{n-2}}_0dt_{n-1}
\Biggl(f^{(m)}(tx-\sum^{n-1}_{j=1}t_j x_j)\nonumber\\ & & \quad \quad
\quad \quad \quad \quad - f^{(m)}(t(x-x_1)-\sum^{n-1}_{j=1}t_j
x_{j+1})\Biggr), \end{eqnarray}
when $f(x)$ is a convergent power series of $x$.

Proof of Formula 8 : Let $h_{n,m}(t;x, x_1, \cdots, x_n)$ be the left-hand
side of (3.33) minus the right-hand side of (3.33). Then, we have
\begin{equation}
\frac{d}{dt}h_{n,m}(t;x, x_1, \cdots, x_n)=xh_{n,m+1}(t;x, x_1, \cdots,
x_n) +h_{n-1,m}(t;x-x_1, x_2, \cdots, x_n).
\end{equation}
If we assume that $h_{n-1,m}(t;x, x_1, \cdots, x_{n-1})=0$ for all
positive integers $m$ and for any $x$, and $\{x_j\}$, then we obtain
\begin{equation}
\frac{d}{dt}h_{n,m}(t;x, x_1, \cdots, x_n)=xh_{n,m+1}(t;x, x_1, \cdots,
x_n). \end{equation}
Thus, we derive
\begin{equation}
\frac{d^N}{dt^N}h_{n,m}(t;x, x_1, \cdots, x_n)=xh_{n,m+N}(t;x, x_1,
\cdots, x_n)\end{equation} for any positive integer $N$. Thus, when $f(x)$
is a polynomial of $x$, we have $h_{n,m+N}(t;x,x_1, \cdots, x_n)=0$ for a
large $N$. Clearly we have \begin{equation}
\left[\frac{d^k}{dt^k}h_{n,m}(t;x, x_1, \cdots, x_n) \right]_{t=0}=0
\end{equation}
for any non-negative integer $k (\leq N)$. The solution of Eq.(3.36) with
(3.37) is given by
\begin{equation}
h_{n,m}(t;x,x_1, \cdots, x_n)=0
\end{equation}
for any positive integers $n$ and $m$. Therefore, when $f(x)$ is a
convergent power series of $x$, we obtain Formula 8 by mathematical
induction, because both sides of Eq.(3.33) is linear with respect to the
function $f(x)$.

Thus, Theorem IV has been proven. An alternative proof of it is given in
Appendix. The third proof is discussed in Sec.VI.

\vspace*{0.3cm}
\noindent
(ii-3) Operator Taylor expansion and shift-hyperoperator ${\cal S}_A(B)$

Now we study the Taylor expansion of $f(A+xB)$. First we prove the
following general Taylor expansion formula.

\vspace*{0.3cm}
\noindent
{\bf Theorem V} : {\it When} $f(A) \in {\cal D}_A$, {\it we have}
\begin{eqnarray}
f(A+xB)& =& \sum^{\infty}_{n=0} x^n \delta_{A \rightarrow B}^n f(A)=
\sum^{\infty}_{n=0} \frac{x^n}{n !} d_{A \rightarrow B}^n f(A)\nonumber\\
& = & \sum^{\infty}_{n=0} \frac{x^n}{n !} \frac{d^nf(A)}{dA^n}:
\underbrace{B \cdot {\mbox{ - - - }} \cdot B}_n. \end{eqnarray}
{\it Equivalently,}
\begin{equation}
f(A+xB)={\cal S}_A(xB) f(A),
\end{equation}
{\it where the shift-hyperoperator} ${\cal S}_A(B)$ {\it is given by}
\begin{equation}
{\cal S}_A(B) \equiv \sum^{\infty}_{n=0} \frac{1}{n !} (d_{A \rightarrow
B})^n ={\rm e}^{d_{A \rightarrow B}}.
\end{equation}

The proof of this theorem is given as follows. From Eqs.(2.8) and (2.17),
we have \begin{equation}
\delta^n_{A \rightarrow B}A^m=
\{A^{m-n}B^n\}_{{\rm sym}}=
\frac{1}{n !} \left[\frac{d^n}{dx^n}(A+xB)^m \right]_{x=0} \end{equation}
for $m \geq n$, and we have $\delta_{A \rightarrow B}^n A^m=0$ for $m<n$.
Therefore, we obtain
\begin{equation}
\delta_{A \rightarrow B}^nf(A) =\frac{1}{n !}
\left[\frac{d^n}{dx^n}f(A+xB)\right]_{x=0} \end{equation}
for any positive integer $n$, when $f(A) \in {\cal D}_A$. This yields
Theorem V.

In particular, if we put $B=dA$, we obtain the following result.

\vspace*{0.3cm}
\noindent
{\bf Theorem VI} : {\it When} $f(A) \in {\cal D}_A$, {\it we have}
\begin{equation}
f(A+xdA)=f(A)+
\sum^{\infty}_{n=1} \frac{x^n}{n!} d^nf(A)={\rm e}^{xd} f(A) \end{equation}
{\it with the differential hyperoperator d defined by} (3.23) {\it , namely}
\begin{equation}
d \equiv d_{A \rightarrow dA}.
\end{equation}

\vspace*{0.5cm}
\noindent
{\bf IV. Multivariate Quantum Analysis}
\setcounter{section}{4}
\setcounter{equation}{0}

In this section, we formulate multivariate quantum analysis, in which we
consider a set of noncommuting power series $\{f(A_1, \cdots, A_q)\}
\equiv \{f(\{A_k\})\}$. This domain is denoted by ${\cal D}_{\{A_k\}}$,
namely
$f(\{A_k\}) \in {\cal D}_{\{A_k\}}$. If we start from a complex number function
$f(\{x_k\})$, it is a problem how to define the operator function
$f(\{A_k\})$, as is well known in quantum mechanics. Here, we start from
the operator function $f(\{A_k\})$ itself which is specified in some
appropriate procedures such as normal ordering.

A definition of the partial differential $d_jf(\{A_k\})$ corresponding to
Eq.(1.1) is given by \begin{equation}
d_jf=\lim_{h \rightarrow 0} \frac{f(A_1, \cdots, A_j+hdA_j, \cdots,
A_q)-f(\{A_k\})}{h}. \end{equation}
Norm convergence of Eq.(4.1) can be discussed in a Banach space and strong
convergence is appropriate for unbounded operators. An algebraic partial
differential corresponding to Eq.(1.2) is given by \begin{equation}
d_jf(\{A_k\})=[H_j, f(\{A_k\})]
\end{equation}
with some auxiliary operators $\{H_j\}$. Both satisfy the Leibniz rule. In
the present paper, we study general properties of multivariate quantum
derivatives which are invariant for any choice of definitions of
differentials.
This invariance can be easily proved by extending the procedure shown in
IIIA. Namely
we have $d_{j}=\delta _{B_j dA_{j};A_{j}}$ with $B_j = -\delta^{-1}_{A_j}$.
The total
differential $df$ is defined by \begin{equation}
df=\sum^q_{j=1} d_jf =(\sum_j d_j)f,
\end{equation}
when $f \in {{\cal D}}_{\{A_k\}}$. The $n$th differential $d^nf$ is also
defined by \begin{equation}
d^nf=(\sum_j d_j)^n f.
\end{equation}
Clearly, $\{d_j\}$ commute with each other, namely $d_jd_k=d_kd_j$, in the
domain ${{\cal D}}_{\{A_k\}}$. One of the key points in the multivariate
quantum analysis is to express $d^nf$ in the form \begin{equation}
d^nf=n! \sum_{j_1, \cdots, j_n} f^{(n)}_{j_1, \cdots, j_n} : dA_{j_1}
\cdot \mbox{- - -} \cdot dA_{j_n}.
\end{equation}
Then, we study how to calculate the hyperoperators $\{f^{(n)}_{j_1,
\cdots, j_n}\}$ in Eq.(4.5).

\vspace*{0.3cm}
\noindent
(i) Ordered differential hyperoperator

In order to study $f^{(n)}_{j_1, \cdots, j_n}$, we introduce here an
ordered differential hyperoperator $d_{j_1, j_2, \cdots, j_n}$ as follows
: \begin{equation}
d_{j_1, j_2, \cdots, j_n}=(d_{j_1}d_{j_2} \cdots d_{j_n})_{{\rm ordered}},
\end{equation}
which means
$d_{j_1, j_2, \cdots, j_n} f(\{A_k\})$ is given by those terms (found via
the Leibniz rule) of $d_{j_1}d_{j_2} \cdots d_{j_n} f(\{A_k\})$ in which
the differentials appear in the order $dA_{j_1}dA_{j_2} \cdots dA_{j_n}$.

For example, we consider an operator function $f(A,B)=ABA^2$. Then we have
\begin{eqnarray}
&d_{A, B}f&=(dA)(dB)A^2, \; \; d_{B, A}f=A(dB)(dA^2)=A(dB)[(dA)A+AdA],
\nonumber\\ &d_{A, A}f&=(dA)BdA^2, \; \; d_{B, B}f=0. \end{eqnarray}

Thus, using this ordered differential, we obtain the following formula.

\vspace*{0.3cm}
\noindent
{\bf Formula 9} : In the domain ${{\cal D}}_{\{A_k\}}$, we have
\begin{equation}
d_{j_1} \cdots d_{j_n}=\sum_{{\rm P}} d_{j_1, \cdots, j_n}. \end{equation}
Here, $\Sigma_{{\rm P}}$ denotes the summation all over the permutations
of $(j_1, \cdots, j_n)$.

The proof will be self-evident.
In particular, we have the following formulas.

\vspace*{0.3cm}
\noindent
{\bf Formula 10} : In the domain ${{\cal D}}_{\{A_k\}}$, we have
$d_jd_k=d_{j,k}+d_{k,j}$ and
\begin{equation}
d_j^n=n! \underbrace{d_{j, \cdots, j}}_n. \end{equation}

\vspace*{0.3cm}
\noindent
{\bf Formula 11} : In the domain ${{\cal D}}_{\{A_k\}}$, we have
\begin{equation}
d^nf=n! \sum_{j_1, \cdots, j_n} d_{j_1, \cdots, j_n} f \end{equation}
for any positive integer $n$.

Under these preparations, we find a procedure to calculate
$\{f^{(n)}_{j_1, \cdots, j_n}\}$ in Eq.(4.5). In principle, they are
obtained through the following relation : \begin{equation}
f^{(n)}_{j_1, \cdots, j_n} :
dA_{j_1} \cdot \mbox{- - -} \cdot dA_{j_n}= d_{j_1, \cdots, j_n} f.
\end{equation}
Here, $d_{j_1, \cdots, j_n}f$ is expressed in the form
\begin{equation}
d_{j_1, \cdots, j_n} f=\sum_k f_{k,0} (dA_{j_1})f_{k,1} (dA_{j_2})
f_{k,2} \cdots (dA_{j_n}) f_{k,n}
\end{equation}
with some appropriate operators $\{f_{k,j}\}$. In order to find
$\{f^{(n)}_{j_1, \cdots, j_n}\}$ explicitly, we have to rearrange
Eq.(4.12) in the form of the left-hand side of Eq.(4.11). For this
purpose, the following rearrangement formula$^{13}$ will be useful.

\vspace*{0.3cm}
\noindent
{\bf Formula 12 (Rearrangement formula)} : Any product $Q_1f_1Q_2f_2
\cdots Q_nf_n$ can be rearranged in the form \begin{equation}
Q_1f_1 \cdots Q_nf_n=\sum^{n+1}_{j=1}f_1f_2 \cdots f_{j-1} \sum_{\pi}
\partial_{\pi(j,j_1)}\partial_{\pi(j_1,j_2)} \cdots
\partial_{\pi(j_k,n+1)} : Q_1 \cdot \mbox{- - -} \cdot Q_n
\end{equation}
with $f_0=1$. Here, $\Sigma_{\pi}$ denotes the summation all over the
ways of the following division of the set of natural numbers $(j, j +
1, \cdots, n-1, n)$ : \begin{equation}
(j, j + 1, \cdots, n-1, n)=\pi(j,j_1)\pi(j_1,j_2) \cdots
\pi(j_k,n+1), \end{equation}
and
\begin{equation}
\pi(j,k)=(j, j+1, \cdots, k-1)
\end{equation}
with $j<j_1< \cdots < j_k \leq n$. Furthermore, the hyperoperator
$\partial_{\pi(j,k)}$ is defined by
\begin{equation}
\partial_{\pi (j,k)} = -\delta_{f_j f_{j+1} \cdots f_{k-1} ; Q_j },
\end{equation}
using the partial inner derivation
$\delta_{f ; Q_j} \equiv \delta_{f ; j }$ which operates only on
$Q_j$ in (4.13).

The proof of Formula 12 is easily given by mathematical induction.

It will be instructive to give here some examples : \begin{eqnarray}
&\;& Q_1f_1=(f_1-\delta_{f_1}) : Q_1,\nonumber\\ &\;&
Q_1f_1Q_2f_2=(f_1f_2-f_1\delta_{f_2;2} -\delta_{f_1f_2;1}+
\delta_{f_1;1}\delta_{f_2;2}) : Q_1 \cdot Q_2,\nonumber\\ &\;&
Q_1f_1Q_2f_2Q_3f_3=
(f_1f_2f_3- f_1f_2\delta_{f_3;3}+
f_1\delta_{f_2;2}\delta_{f_3;3}-f_1\delta_{f_2f_3;2}\nonumber\\ &\;&
-\delta_{f_1f_2f_3;1 }
+ \delta_{f_1;1}\delta_{f_2 f_3;2}+ \delta_{f_1f_2;1} \delta_{f_3;3}
-\delta_{f_1;1}\delta_{f_2;2} \delta_{f_3;3}) : Q_1 \cdot Q_2 \cdot
Q_3. \end{eqnarray}

\vspace*{0.3cm}
\noindent
(ii) Partial derivative and multivariate
operator Taylor expansion

It will be convenient to define the following partial quantum
derivative \begin{equation}
\frac{\partial ^n f}{\partial A_{j_n} \cdots \partial A_{j_1}} \equiv
n ! f^{(n)}_{j_1, \cdots, j_n},
\end{equation}
using the hyperoperators $\{f^{(n)}_{j_1, \cdots, j_n}\}$ determined
through the relation (4.11). Then, we obtain the following theorem.

\vspace*{0.3cm}
\noindent
{\bf Theorem VII} : {\it When} $f(\{A_k\}) \in {{\cal D}}_{\{A_k\}}$,
{\it we have } \begin{eqnarray}
f(\{A_j+x_jdA_j\}) & = & \sum^{\infty}_{n=0} \sum_{j_1, \cdots, j_n}
x_{j_1} \cdots x_{j_n} d_{j_1, \cdots, j_n}f\nonumber\\ & = &
\sum^{\infty}_{n=0} \sum_{j_1, \cdots, j_n} x_{j_1} \cdots x_{j_n}
f^{(n)}_{j_1, \cdots, j_n} : dA_{j_1} \cdots dA_{j_n}\nonumber\\ & =
& \sum^{\infty}_{n=0} \sum_{j_1, \cdots, j_n} \frac{x_{j_1} \cdots
x_{j_n}}{n !}
\frac{\partial ^n f}{\partial A_{j_n} \cdots \partial A_{j_1}} :
dA_{j_1} \cdots dA_{j_n}.
\end{eqnarray}
{\it Equivalently, we have}
\begin{equation}
f(\{A_j+x_j B_j\})={\rm exp}\left(\sum^{q}_{j=1} x_j d_{A_j
\rightarrow B_j}\right)f(\{A_j\}). \end{equation}
{\it In particular,}
\begin{equation}
f(\{A_j+xdA_j\})={\rm e}^{xd}f(\{A_j\})=
\sum^{\infty}_{n=0}\frac{x^n}{n !}d^nf(\{A_j\}) \end{equation}
{\it with } $d=\Sigma_j d_j$.

\vspace*{0.5cm}
\noindent
{\bf V. Auxiliary Operator Method }
\setcounter{section}{5}
\setcounter{equation}{0}

It will be convenient to introduce the auxiliary operators $\{H_j\}$
satisfying the following conditions: \begin{equation}
[H_j, H_k]=0, \; [H_j, A_k]=0, \; [H_j, [H_k, A_k]]=0, \; \; {{\rm
for}} \; \; j \neq k \end{equation}
and
\begin{equation}
[H_j, [H_j, A_j]]=0.
\end{equation}
Using these auxiliary operators $\{H_j\}$, we introduce the following
partial differential \begin{equation}
d_jf \equiv [H_j,f] \equiv \delta_{H_j}f. \end{equation}
In particular, we have
\begin{equation}
dA_j=d_jA_j=[H_j, A_j],
\end{equation}
and
\begin{equation}
d^2A_j=0 \; \; {\rm and} \; \; d_j(dA_k)=0. \end{equation}
The total differential $df$ is given by
\begin{equation}
df \equiv \sum_j [H_j, f]=(\sum_j d_j)f. \end{equation}

One of the merits of this auxiliary operator method is that we can
easily obtain the operator Taylor expansion as follows :
\begin{eqnarray}
\lefteqn{{\rm exp}(\sum_j x_j \delta_{H_j})f(\{A_j\}) = f(\{{\rm
exp}(x_j\delta_{H_j})A_j\})}\nonumber\\ & & = f(\{A_j+
x_j\delta_{H_j} A_j\}) = f(\{A_j+ x_j dA_j\}), \end{eqnarray}
using Eqs. (5.2) and (5.4). That is, we have \begin{equation}
f(\{A_j+ x_jdA_j\})={\rm exp}(\sum_j x_j \delta_{H_j})f(\{A_j\})
={\rm e}^{\Sigma_j x_j d_j} f(\{A_j\}),
\end{equation}
using the relation (5.3).

\vspace*{0.5cm}
\noindent
{\bf VI. Some General Remarks and Applications to Exponential Product
Formulas}\setcounter{section}{6} \setcounter{equation}{0}

It will be instructive to remark that when the operator $A$ depends
on a parameter $t$, namely $A=A(t)$, we have$^{10}$ \begin{equation}
\frac{df(A(t))}{dt}=\frac{df(A(t))}{dA(t)} \cdot \frac{dA(t)}{dt}.
\end{equation}
This formula insures again the invariance of the derivative
$df(A(t))/dA$, because $df(A(t))/dt$ and $dA(t)/dt$ do not depend on
the choice of the differential $df(A(t))$. Furthermore, we have
\begin{equation}
\frac{df(g(A))}{dA}=\frac{df(g(A))}{dg(A)} \cdot \frac{dg(A)}{dA},
\end{equation}
because
\begin{equation}
df(g(A))=\frac{df(g(A))}{dg(A)} \cdot dg(A) =\frac{df(g(A))}{dg(A)}
\cdot \frac{dg(A)}{dA} : dA. \end{equation}

It will be also interesting to note the derivative of hyperoperators.
The first differential of a hyperoperator $f(\delta_A)$ is given
by$^{10}$
\begin{equation}
d(f(\delta_A)dA)=
\int^1_0 dt[f^{(1)} (t \delta_1+\delta_2)- f^{(1)} (\delta_1+t
\delta_2)] : (dA)^2. \end{equation}
In general, we have
\begin{eqnarray}
\lefteqn{d[f(A ; \delta_1, \cdots, \delta_n) : (dA)^n]= \int^1_0
dtf^{(1)} (A-t \delta_1 ; \delta_2, \cdots, \delta_{n+1}) :
(dA)^{n+1}}\nonumber\\ & & + \sum^n_{k=1}\{
\int^1_0 dt_k \Biggl[f^{(1,k)} (A ; \delta_1, \cdots, \delta_{k-1},
t_k \delta_k +\delta_{k+1}, \delta_{k+2}, \cdots,
\delta_{n+1})\nonumber\\ & & -f^{(1,k)} (A ; \delta_1, \cdots,
\delta_{k-1},  \delta_k + t_{k} \delta_{k+1}, \delta_{k+2}, \cdots,
\delta_{n+1})\Biggr] :(dA)^{n+1}\}. \end{eqnarray}
Here, $f^{(1)}(x ; x_1 \cdots, x_n)$ denotes the first derivative of
$f(x ; x_1 \cdots, x_n)$ with respect to $x$ and $f^{(1,k)}(x ; x_1
\cdots, x_k, \cdots , x_n)$ denotes the first derivative of $f$ with
respect to $x_k$. Note that $A$ and $\{\delta_k\}$ commute with each
other. These formulas will be also useful in proving Theorem IV. In
fact, we obtain \begin{eqnarray}
\lefteqn{d^2f(A)=d(df(A))}\nonumber\\
& & =d(\int^1_0 dt f^{(1)}(A-t \delta_A)dA)\nonumber\\ & & =\int^1_0
dt d(f^{(1)}(A-t \delta_A) \cdot dA)\nonumber\\ & & =\int^1_0 dt_1
\int^1_0 dt_2 f^{(2)}(A-t_1 \delta_1 -t_2 \delta_2) :
(dA)^2\nonumber\\ & & + \int^1_0 dt \int^1_0 (-s) ds
[f^{(2)}(A-s(t \delta_1 +\delta_2))-
f^{(2)}(A-s(\delta_1 +t\delta_2))] : (dA)^2\nonumber\\ & & =2
\int^1_0 dt_1 \int^{t_1}_0 dt_2 f^{(2)}(A-t_1 \delta_1 -t_2 \delta_2)
: (dA)^2. \end{eqnarray}
Similarly we can derive Theorem IV using the above formula (6.5).

There are many applications of quantum analysis to exponential
product formulas$^{11-15}$ such as the Baker-Campbell-Hausdorff
formula.

For example, if we put
\begin{equation}
{\rm e}^{A_1(x)}{\rm e}^{A_2(x)} \cdots {\rm e}^{A_r(x)}={\rm
e}^{\Phi(x)}, \end{equation}
the operator $\Phi(x)$ is shown to satisfy the operator
equation$^{14}$ \begin{equation}
\frac{d \Phi(x)}{dx} = \Delta^{-1} (\Phi(x)) \sum^r_{j=1} {\rm
exp}(\delta_{A_1(x)}) \cdots {\rm exp}(\delta_{A_{j-1}(x)})
\Delta(A_j(x)) \frac{dA_j(x)}{dx} \end{equation}
using the quantum derivative of ${\rm e}^{A}$ : \begin{equation}
\frac{d{\rm e}^{A}}{dA}=
\frac{{\rm e}^{A}-{\rm e}^{A-\delta_A}}{\delta_A} ={\rm
e}^{A}\Delta(-A) ;
\Delta(A)=\frac{{\rm e}^{\delta_A}-1}{\delta_A}. \end{equation}
The solution of Eq.(6.8) is given by
\begin{eqnarray}
\Phi(x) & = & \sum^r_{j=1}\int^x_0
\frac{\log[{\rm exp}(\delta_{A_1(t)}) \cdots {\rm
exp}(\delta_{A_r(t)})]} {{\rm exp}(\delta_{A_1(t)}) \cdots {\rm
exp}(\delta_{A_r(t)})-1}\nonumber\\ & \times & {\rm
exp}(\delta_{A_1(t)}) \cdots {\rm exp}(\delta_{A_{j-1}(t)})
\Delta(A_j(t)) \frac{dA_j(t)}{dt} dt+ \Phi(0). \end{eqnarray}
This is a generalized BCH formula.

In particular, we have
\begin{equation}
\log({\rm e}^A {\rm e}^B{\rm e}^A)=
\int^1_0 \left(\frac{{\rm e}^{t \delta_A} {\rm e}^{\delta_B} {\rm
e}^{t \delta_A}+1} {{\rm e}^{t \delta_A} {\rm e}^{\delta_B}{\rm e}^{t
\delta_A}-1} \log ({\rm e}^{t \delta_A} {\rm e}^{\delta_B} {\rm e}^{t
\delta_A})\right) A dt+B.
\end{equation}
Recursively we have
\begin{equation}
\log ({\rm e}^{A_1} \cdots {\rm e}^{A_r})= \int^1_0 dt \frac{\log
E_r(t)}{E_r(t)-1} (A_1+E_r(t)A_r)+ \Phi_{2,r-1}, \end{equation}
where $\Phi_{2, r-1} =\log ({\rm e}^{A_2} \cdots {\rm e}^{A_{r-1}})$,
and \begin{equation}
E_r(t)={\rm exp}(t \delta_{A_1}) {\rm exp}(\delta_{A_2}) \cdots
{\rm exp}(\delta_{A_{r-1}}){\rm exp}(t \delta_{A_r}). \end{equation}

The feature of these formulas is that $\Phi(x)$ and $\log({\rm
e}^{A_1} \cdots {\rm e}^{A_r})$ are expressed only in terms of linear
combinations of $\{A_j\}$ and their commutators.

These formulas will be useful in studying higher-order decomposition
formulas$^{17}$.

\vspace*{0.5cm}
\noindent
{\bf VII. Summary and Discussion}

In the present paper, we have unified an analytic formulation of
quantum analysis based on the G\^ateaux differential and an algebraic
formulation of quantum analysis based on commutators, by introducing
the two hyperoperators $\delta_{A \rightarrow B} \equiv
-\delta_A^{-1}\delta_B$ and $d_{A \rightarrow B} \equiv
\delta_{(-\delta_A^{-1} B) ; A}$. This general theory of quantum
analysis gives a proof of the invariance of quantum derivatives for
any choice of the definitions of differentials in the domain ${\cal
D}_A$. This domain can be easily extended$^{12}$ to the region
$\tilde{{\cal D}}_A$ which is a set of convergent Laurent series of
the operator $A$ in a Banach space.
Multivariate quantum derivatives have also been formulated using the
rearrangement formula.

The present general formulation will be used effectively in studying
quantum fluctuations in condensed matter physics and it will be also
useful in mathematical physics. The present quantum analysis can also
be extended to an infinite number of variables$^{14}$. The quantum
analysis has been also used$^{15}$ in extending Kubo's linear response
theory$^{18}$ and Zubarev's theory of statistical operator$^{19}$ to
more general nonlinear situations$^{11}$. The invariant property of
quantum derivatives derived in Sec.III is closely related$^{15}$ to the
general feauture of the fluctuation-dissipation theorem$^{18-21}$.
General quantum correlation identities are also derived$^{15}$ using the
quantum analysis. For the convergence of unbounded operators, see the second
paper of Ref.17.

\vspace*{0.5cm}
\noindent
{\bf Acknowledgements}

The author would like to thank Prof. K. Aomoto, Prof. H. Araki and
Prof. H. Komatsu for useful discussion at the Hakone Meeting, and
also thank Dr. H.L. Richards for a kind reading of the manuscript.
The referee's comments have been very helpful to improve the manuscript.
The author would also like to thank Noriko Suzuki for continual
encouragement.

This study is partially financed by the Research Fund of the Ministry
of Education, Culture and Science.

\vspace*{0.5cm}
\begin{center}
{\bf Appendix } : Alternative Proof of Theorem IV \end{center}
\renewcommand{\theequation}
{A.\arabic{equation}}
\setcounter{equation}{0}

First we study the case $f(A)=A^m$ for a positive integer $m$. The
$n$th differential $d^nA^m$ is expressed in the form \begin{eqnarray}
d^nA^m &= & d^n_{A \rightarrow dA} A^m=n ! \delta^n_{A \rightarrow
dA}A^m =n! \{A^{m-n}(dA)^n\}_{{\rm sym}}\nonumber\\ & = & n !
\sum_{k_j \geq 0, \sum k_j=m-n} A^{k_0}(dA)A^{k_1}(dA) \cdots
A^{k_{n-1}}(dA)A^{k_n}\nonumber\\ & = & n ! \sum_{k_j \geq 0, \sum
k_j=m-n} A^{k_0}(A-\delta_1)^{k_1}\nonumber\\
& \cdots &
(A-\delta_1 - \cdots -\delta_n)^{k_n} : dA \cdot {\mbox{- - -}} \cdot
dA, \end{eqnarray}
for $m \geq n$ and $d^nA^m=0$ for $n>m$, using Theorem III, Formula
5, Formula 2, the definition of the symmetrized product, Eq.(2.7),
and the following formula$^{10}$.

\vspace*{0.3cm}
\noindent
{\bf Formula A} : For any operator $Q$, we have \begin{equation}
Qf(A)=f(A-\delta_A)Q
\end{equation}
when $f(A) \in {\cal D}_A$.

This yields Lemma 2. Now, we prove the following lemma.

\vspace*{0.3cm}
\noindent
{\bf Lemma A} : When $f(A)=A^m$ with a positive integer $m ( \geq
n)$, the formula (3.28) holds. That is, we have \begin{equation}
\sum_{k_j \geq 0, \sum k_j=m-n} A^{k_0}
(A-\delta_1)^{k_1} \cdots (A-\delta_1- \cdots -\delta_n)^{k_n}=F_n(A;
\delta_1 \cdots, \delta_n), \end{equation}
where
\begin{equation}
F_n(A; \delta_1, \cdots, \delta_n) \equiv \frac{m!}{(m-n)!}
\int^1_0dt_1 \int^{t_1}_0 dt_2 \cdots \int^{t_{n-1}}_0dt_n
(A-\sum^n_{j=1} t_j\delta_j)^{m-n}. \end{equation}

This lemma can be proved by mathematical induction as follows. We
assume that Eq.(A.3) holds in the case of $F_{n-1}(A ; \delta_1
\cdots, \delta_{n-1})$. Then, we have \begin{eqnarray}
\lefteqn{F_n(A; \delta_1, \cdots, \delta_n) = \frac{m!}{(m-n+1)!}
\int^1_0dt_1 \int^{t_1}_0 dt_2 \cdots
\int^{t_{n-2}}_0dt_{n-1}}\nonumber\\
& & \times
\frac{1}{\delta_n}\left[(A-\sum^{n-1}_{j=1}t_j\delta_j)^{m-n+1} -
\{A-\sum^{n-2}_{j=1}t_j\delta_j
-t_{n-1}(\delta_{n-1}+\delta_n)\}^{m-n+1}\right]\nonumber\\ & & =
\frac{1}{\delta_n} \sum_{k_j \geq 0, \sum k_j=m-n+1}
A^{k_0}(A-\delta_1)^{k_1} \cdots (A-\delta_1- \cdots
-\delta_{n-2})^{k_{n-2}}\nonumber\\ & & \times \{(A -\delta_1- \cdots
-\delta_{n-1})^{k_{n-1}} - (A -\delta_1- \cdots
-\delta_n)^{k_{n-1}}\}\nonumber\\ \end{eqnarray}
under the assumption that Eq.(A.3) holds for $F_{n-1}(A; \delta_1,
\cdots, \delta_{n-1})$. Then, the above expression (A.5) can be
rearranged as \begin{eqnarray}
\lefteqn{F_n(A; \delta_1, \cdots, \delta_n) = \sum_{k_j \geq 0, \sum
k_j=m-n+1}
A^{k_0}(A-\delta_1)^{k_1} \cdots (A-\delta_1- \cdots
-\delta_{n-2})^{k_{n-2}}}\nonumber\\ & & \times
\sum_{k'_{n-1} \geq 0, k'_n \geq 0, k'_{n-1}+k'_n= k_{n-1}-1} (A
-\delta_1- \cdots -\delta_{n-1})^{k'_{n-1}} \cdot (A -\delta_1- \cdots
-\delta_n)^{k'_n}\nonumber\\ & & = \sum_{k_j \geq 0, \sum k_j=m-n}
A^{k_0} (A -\delta_1)^{k_1} \cdots
(A -\delta_1- \cdots -\delta_n)^{k_n},
\end{eqnarray}
by noting that $k_0+k_1+ \cdots +k_{n-2} +k'_{n-1}+k'_n=k_0+k_1+
\cdots + k_{n-1} -1 = (m-n+1)-1=m-n$. Thus, we arrive at Lemma A. Any
operator $f(A) \in {\cal D}_A$ is expressed as a power series of
$\{A^m\}$. Then, Lemma A yields Theorem IV.

\vspace*{0.5cm}
\noindent
{\bf Rerferences}
\baselineskip=0.2cm
\begin{enumerate}
\item E. Hille and R.S. Phillips, {\it Functional analysis and
semi-groups,} Amer. Soc. Math. Colloq. Publ. {\bf 31} (1957).

\item L. Nachbin, {\it Topology on Spaces of Holomorphic Mappings, }
(Springer-Verlag, 1969).

\item W. Rudin, Functional Analysis (McGraw Hill, 1973).

\item M.C. Joshi and R.K. Bose, {\it Some topics in nonlinear
functional analysis,} (Wiley 1985).

\item K. Deimling, {\it Non-linear functional analysis} (Springer,
1985).

\item S. Sakai, {\it Operator Algebra in Dynamical Systems,}
Cambridge Univ. Press (1991).

\item M.V. Karasev and V.P. Maslov, {\it Nonlinear Poisson Brackets}
--- {\it Geometry and Quantization} (Translations of Mathematical
Monographs, Vol.119, Am. Math. Soc. 1993).

\item A. Connes, {\it Noncommutative Geometry} (Academic Press, Inc.
1994).

\item V.E. Nazaikinskii, V.E. Shatalov and B.Yu. Sternin, {\it
Methods of Noncommutative Analysis} (Walter de Gruter, 1996).

\item M. Suzuki, {\it Quantum Analysis } --- {\it Noncommutative
differential and integral calculi,} Commun. Math. Phys. {\bf 183},
339 (1997) .

\item M. Suzuki, Int. J. Mod. Phys. {\bf B10}, 1637 (1996).

\item M. Suzuki, Phys. Lett. {\bf A224}, 337 (1997).

\item M. Suzuki, Trans. of J. Soc. for Ind. and Appl.
Math. (in Japanese), Vol.7, No.3, 257 (1997).

\item M. Suzuki, J. Math. Phys. {\bf 38}, 1183 (1997).

\item M. Suzuki, submitted to Int. J. Mod. Phys. {\bf B}.
See also Sec.X.4 of R. Bhatia, {\it Matrix Analysis}
(Springer, 1997).

\item M. Abe, N. Ikeda and N. Nakanishi, {\it Operator ordering index
method for multiple commutators and Suzuki's quantum analysis}
(preprint).

\item M. Suzuki, Commun. Math. Phys. {\bf 163}, 491 (1994), and
references cited therein. See also M. Suzuki, Rev. of Math. Phys.
{\bf 8}, 487 (1996).

\item R. Kubo, J. Phys. Soc. Jpn. {\bf 12}, 570 (1957).

\item D.N. Zubarev, {\it Nonequilibrium Statistical Mechanics}
(Nauka, 1971).

\item R. Kubo, M. Yokota and S. Nakajima, J. Phys. Soc. Jpn. {\bf
12}, 1203 (1957).

\end{enumerate}
\end{document}